\documentclass[11pt,preprint]{aastex}

\usepackage{graphicx}
\usepackage{color}
\usepackage{amsmath}
\usepackage{natbib}
\advance \voffset by 0.00cm\relax

\begin{document}

\title{A Possible Period for the K-band Brightening Episodes of GX 17+2}

\author{
	Jillian Bornak\altaffilmark{1}\altaffilmark{\dag}, 
	Bernard J. McNamara\altaffilmark{1}\altaffilmark{\dag}\altaffilmark{\ddag},
	Thomas E. Harrison\altaffilmark{1}\altaffilmark{\dag}, 
	Michael P. Rupen\altaffilmark{2},
	Reba M. Bandyopadhyay\altaffilmark{3}\altaffilmark{\ddag},
	Stefanie Wachter\altaffilmark{4}
	}

\altaffiltext{1}{Department of Astronomy, New Mexico State University,
	1320 Frenger Mall, Las Cruces, NM  88003; jbornak@nmsu.edu, bmcnamar@nmsu.edu, tharriso@nmsu.edu}
\altaffiltext{2}{NRAO,
	1003 Lopezville Road,  Socorro, NM 87801; mrupen@aoc.nrao.edu}
\altaffiltext{3}{Department of Astronomy, University of Florida,
	Gainesville, FL 32611; reba@astro.ufl.edu}
\altaffiltext{4}{{\it Spitzer} Science Center, Caltech M/S 220-6, 1200 E. California Blvd., Pasadena, CA 91125; wachter@ipac.caltech.edu}
\altaffiltext{\dag}{Visiting astronomer, Kitt Peak National Observatory, National Optical Astronomy Observatory, which is operated by the Association of Universities for Research in Astronomy (AURA) under cooperative agreement with the National Science Foundation}
\altaffiltext{\ddag}{Visiting astronomer, Cerro Tololo Inter-American Observatory, National Optical Astronomy Observatory, which is operated by the Association of Universities for Research in Astronomy (AURA) under cooperative agreement with the National Science Foundation}

\begin{abstract} 
The low mass X-ray binary and Z source GX 17+2 undergoes infrared {\it K}-band brightening episodes of at least 3.5 magnitudes. The source of these episodes is not known. Prior published {\it K}-band magnitudes and new {\it K}-band measurements acquired between 2006 and 2008 suggest that the
episodes last at least 4 hours and have a period of 3.01254 $\pm$ 0.00002 days.  Future bright episodes can be predicted using the ephemeris $JD_{max}(n) = 2454550.79829 + (3.01254 \pm 0.00002$)(n) days. A growing body of evidence suggests that the GX 17+2 could have a synchrotron jet, which could cause this activity. \\
\end{abstract}


\section{Introduction}
The low mass X-ray binary (LMXB) GX 17+2 is one of the brightest persistent X-ray sources in the sky and its high energy behavior has been extensively monitored by X-ray satellites.  It is one of only eight objects known as Z sources, so-called for the track they trace out in an X-ray color-color diagram.  Z sources are thought to be neutron star binaries accreting near or at the Eddington rate, and position along the Z is thought to reflect a change in mass accretion rate.  The radio emission of GX 17+2 is already known to be connected with position in this diagram \citep{Penninx, Migliari}. Associating the multiwavelength behavior of a LMXB with its Z position provides an excellent way in which to test theoretical models of these systems \citep{Psaltis}. 

In contrast to the well-studied X-ray emission from GX 17+2, relatively little is known about this object's optical/infrared counterpart \citep{Davidsen, Naylor, Deutsch}.  It was only recently that \cite{Callanan} determined the IR counterpart of GX 17+2 is blended with a G type field star 0.9" away, which was erroneously given the variable star name NP Ser.  These authors also observed a remarkable 3.5 mag difference between their two {\it K}-band observations.  The underlying cause of these bright episodes remains a mystery.  The X-ray data of GX 17+2, extending back to the early 1980s, do not show any evidence for eclipsing or dipping behavior, so these phenomena cannot be responsible for the variable IR emission \citep{Kuulkers97, Penninx, Homan}. The IR and soft X-ray fluxes also do not appear to be correlated, making it unlikely that they arise from the reprocessing of X-rays in an accretion disk surrounding the system's neutron star or on the surface of the donor star \citep{Bandyopadhyay}.  Finally, free-free emission from an X-ray driven wind would also produce a correlation between the X-ray and IR emission; therefore, this process appears to be ruled out as well \citep{Callanan}.

Some attempts have been made to explain the perplexing nature of the GX 17+2 {\it K}-band 
bright episodes, but they are based on very few measurements.  \cite{Bandyopadhyay} used the Ohio State Infrared IMager/Spectrometer on the Perkins 1.8m telescope at the Lowell Observatory
to obtain eighteen {\it K}-band magnitudes of the blended image of GX 17+2/NP Ser between July 16 and August 20, 1997. These magnitudes were compared to RXTE All-Sky-Monitor (ASM) one-day average X-ray flux measurements to determine whether they are connected, but no correlations were found. These {\it K}-band magnitudes and several hundred days of ASM data were then used 
to conduct a search for periods between 4 and 200 days. No periodic behavior was found.

The study by \citep{Callanan} used four {\it K}-band measurements, only one of which was obtained during an IR bright episode.  \cite{Callanan} argue the case for a jet interpretation for GX 17+2. They note that the levels of the radio and {\it K}-band variability in GX 17+2 are comparable suggesting that they have a common origin and that, if an optically thick synchrotron power law index is assumed, an extrapolation of the maximum GX 17+2 radio flux density predicts a peak magnitude of {\it K} = 14. This is close to the observed value that GX 17+2 reaches during an IR bright state.  Additionally, \cite{Russell} supports the idea that the GX 17+2 {\it K}-band bright episodes are related to the presence of a jet.  Based on an examination of 19 low magnetic field neutron star binaries, they conclude that jets are necessary to explain the quasi-simultaneous optical/IR/X-ray emission from high luminosity systems such as GX 17+2.

The fact that GX 17+2 is a Z-source might allow studies of its X-ray/IR behavior to further our understanding of the multi-wavelength emission processes operating in other peculiar LMXBs.  The unusual nature of the IR emission from GX 17+2 and its very limited observational data base led us to conduct a multi-year campaign to better quantify its {\it K}-band properties.  In this Letter we present the first evidence that the IR bright episodes of GX 17+2 last at least 4 hours and repeat with a period of 3.01245 days.  We also find that the shape of this system's {\it K}-band light curve and its peak brightness are highly variable.

\section{Observations and Data Reduction}
This study is based on new {\it K}-band measurements of GX 17+2 acquired between July 2006 and July 2009 as listed in Table 2.  Measurements were obtained at the Astrophysical Research Consortium's Apache Peak Observatory (APO), and at the National Optical Astronomy Observatory's Kitt Peak National Observatory (KPNO).  Additionally, we had 102 nights of unpublished observations at the Cerro Tololo Inter-American Observatory (CTIO).  The {\it K}-band photometry of GX 17+2 reported by \cite{Callanan} and \cite{Bandyopadhyay} were also used, as well as a {\it Spitzer} MIPS 24 $\mu$m observation \citep{Wachter}.

{\it K}-band photometry was obtained in 2006 on the nights of UT July 1, 12, 14 and in 2009 on the night of UT July 1 using the Near-Infrared Camera and Fabry-Perot Spectrometer (NICFPS)\footnote{http://www.apo.nmsu.edu/arc35m/Instruments/NICFPS/} on the APO 3.5 meter telescope.  NICFPS has a 0.273 arcsec/pixel scale, giving a 4.58 arcmin square field of view.  Exposures were 8s and a five point dither pattern was used.  While NICFPS is capable of resolving GX 17+2 from NP Ser, poor seeing prevented us from doing so.  On October 4, 2007 GX 17+2 was observed  using the Cornell Massachusetts Slit Spectrograph (CorMASS) in slit view mode.  CorMASS is a 256x256 pixel detector with a scale of 0.375 arcsec/pixel in slit mode giving a 1.6 arcmin square field of view.

GX 17+2 was observed in 2007 from May 25 to 30 and in 2008 on July 8 and 13 with the Simultaneous Quad Infrared Imaging Device (SQIID)\footnote{http://www.noao.edu/kpno/sqiid/} on the KPNO 2.1m telescope. SQIID simultaneously collects data in the {\it JHK} filters. SQIID has a scale of 0.69 arcsec/pixel providing a roughly 5 arcmin square field of view. On these images GX 17+2 is blended with the image of NP Ser.  The data were taken with 8s exposures using a two point dither co-adding six images per dither position for 2006 and a five point dither co-adding eight images per dither position for 2007.

The above datasets were analyzed using standard IRAF\footnote{IRAF is distributed by the NOAO, which is operated by AURA under cooperative agreement with the NSF.} reduction routines.  Aperture photometry was performed on the blend of GX 17+2 and NP Ser and relative photometry was performed with respect to four bright field stars, whose {\it K}-band magnitudes were taken from 2MASS Point Source Catalog.  Dome flats were used for flat fields and the dithered images were median combined to create sky images. Flat-fielded and sky subtracted images for each position in the dither were then combined to increase the S/N.  Except for the extra care needed to select uncontaminated calibration stars in the crowded region of GX 17+2, no unusual reduction issues were encountered.

CTIO {\it K}-band measurements of GX 17+2 were obtained during 51 nights in 2001 extending from March 1 to June 9 and for 51 nights in 2007 from March 15 to July 11 with A Novel Dual Imaging CAMera (ANDICAM)\footnote{http://www.astronomy.ohio-state.edu/ANDICAM/detectors.html} on the 1.3m telescope operated by the Small and Moderate Aperture Research Telescope System (SMARTS) consortium.  ANDICAM has a 0.369 arcsec/pixel scale giving a 6.29 arcmin square field of view.  The 2001 data were taken with 30s exposures and a seven point dither co-adding two images per dither position, and the 2007 have 60s exposures and a five point dither pattern.

The reduction of the SMARTS data sets was considerably more challenging. The {\it K}-band images possess a variable fringing pattern that cannot be totally removed. Therefore, although the formal {\it K}-band measurement errors were $\pm0.07$ mag, the transient noise pattern problem produced unreliable photometry for sources as faint as NP Ser and GX 17+2.  Therefore, these data were not included in our period-searching analysis.

Unfortunately, the position of GX 17+2 along its Z was unknown during the IR bright observations.  X-ray data was taken an hour after the 2006 July 12 observation with the Rossi X-ray Timing Explorer (RXTE)\footnote{http://heasarc.gsfc.nasa.gov/docs/xte/rxte.html} satellite's Proportional Counter Array (PCA), where GX 17+2 was found to be on the Normal Branch (McNamara et al. 2009, in prep.).  Even though we have simultaneous RXTE ASM data for our observations, we cannot recover the Z position due to the low X-ray flux and the resultant uncertainties.

\section{The {\it K}-band Period Search}
Table 1 lists the six dates when a bright {\it K}-band episode of GX 17+2 was detected.  During four of the observing sessions at the APO and KPNO, light curves extending for approximately 4 hours were acquired, as shown in Figure 1.  The 2007-10-04 entry in Table 1 refers to the APO CorMass detection during which only four short exposures were obtained; the 1999-06-26 entry refers to the IR bright observation by \cite{Callanan}; and the 2008-10-21 entry refers to a {\it Spitzer} MIPS 24 $\mu$m observation \citep{Wachter}.  These three observations are single images rather than full light curves, and so the time of observation is shown rather than the time of peak IR brightness.  The estimated times given in Table 1 for the center of the bright period are much more uncertain than the size of the given heliocentric correction.

The extent of a GX 17+2 bright episode is defined as the time interval during which the GX 17+2 and NP Ser blended image had a {\it K}-band magnitude brighter than 14.5 (see Callanan et al. 2002 for measurements of NP Ser).  An entire bright episode is not covered by any of our light curves, but our longest-duration light curve shows it can last for at least four hours (see Figure 1).  The differing shapes of these light curves pose obvious problems for the determination of any periodicity.  First, the variety in shape does not permit the light curves to be folded as is done to determine the period of an eclipsing system.  Secondly, the maximum {\it K}-band magnitude from one bright episode to another is not a constant. As shown in Table 1, the peak magnitude can vary by at least $\pm$0.28 magnitudes.  These two properties do not allow snap-shot {\it K}-band magnitudes to be associated with a particular phase.

Given the above limitations, the following procedure was used to determine the period of the IR bright episodes of GX 17+2. The KPNO observations of two IR bright light curves (middle two curves shown in Figure 1) were separated by about three days.  Observations prior to and between these two IR bright times showed GX 17+2 persistently faint, implying a three day period.  Therefore, a grid of phase points was constructed for candidate periods between 2.8 days and 3.2 days for the times during July 2006 and May 2007 listed in Table 1.  The \cite{Callanan} observation was not used in this analysis step.  The standard deviation of the five phase points at each epoch was plotted versus period. This is shown in the top panel of Figure 2.  If the system's IR bright episodes are periodic, each of the times in Table 1 should have nearly the same phase.  This would then produce a minimum in this figure.  In a system whose period is days long, 1-2 hour uncertainties in the mid points of the IR bright times will broaden this minimum.  The result of this computation indicates that the IR bright episodes of GX 17+2 have a period of about 3.012 days.

To better define this value, the bright \cite{Callanan} {\it K}-band observation of GX 17+2 was then included (see their Figure 1), and the above procedure was repeated.  A finer grid step in period was used and the searched interval was reduced to 3.0 to 3.1 days. The results of this exercise are shown in Figure 2 (bottom). The minimum dispersion is located at a period of 3.01254 $\pm$ 0.00002 days.  The dates of our observations were constrained by the small window in which GX 17+2 was visible from APO and KPNO.  The fact that our observations were taken nearly the same time each year causes the multi-cusp pattern in the bottom of Figure 2.

The validity of our period was tested in three different ways.  First, to test the robustness of the period, the center of the bright episode times in Table 1 were randomly changed by 1-2 hours.  The period was then recomputed using the above procedure.  Second, the period was calculated using  the time when the system's brightness stopped its rise.  For the first two tests, the derived period changed by less than 0.00002 days.  Lastly, since this period was initially computed using only the July 2006 and May 2007 data, the ephemeris was used to predict IR bright and faint episodes in October 2007, July 2008, and July 2009.  The observations of GX 17+2 in IR bright and IR faint states confirmed our period and eliminated the possibility of any aliasing.  The period calculation was then redone with the additional times included, and we find that this period was consistent with all of our APO and KPNO non-detections.

We also attempted to use the {\it K}-band observations reported by \cite{Bandyopadhyay} as a test.  {\it None} of the SMARTS observations corresponded to an IR bright time, and no definitive bright episodes were detected in that study.  This result agrees with the brightening period suggested in this Letter.  Based on the KPNO, APO, and Keck data the ephemeris of the GX 17+2 IR bright episodes is $JD_{max}(n) = 2454550.79829 + 3.01254(n) \pm 0.00002$ days.

\section{Discussion}
We have presented the first evidence that the {\it K}-band brightening episodes of GX 17+2 are periodic and that its IR light curve and maximum brightness vary from event to event. Although this result is consistent with all of our {\it K}-band data, we would like to emphasize that additional observations are clearly desirable to confirm the nature of these IR bright episodes and to better quantify the structure of the {\it K}-band light curve.  As mentioned in the prior sections, although several mechanisms have been ruled out as the likely source for the IR bright episodes in GX 17+2, their origin remains a mystery.  The data collected for this study is not adequate to answer this question.  However, we offer the following comments.

\cite{Migliari} interpret the correlated radio and X-ray fluctuations of GX 17+2 as suggestive of a compact jet.  These authors showed that the peak radio flux of GX 17+2 at 4.8 and 8.4 GHz varied smoothly with the location in the Z plot.  Maximum fluxes of 3 mJy and 0.4 mJy, respectively, were found when this system was on its horizontal branch.  Contemporaneous RXTE PCA data with our {\it K}-band bright episode of 2006 July 12 suggest that GX 17+2 was on its normal branch (McNamara et al. 2009, in prep.).  VLA observations obtained five hours after the IR peak had fluxes of 7 mJy and 4.4 mJy at 4.8 and 8.4 GHz, respectively.  These are the highest flux levels ever observed for this system at these frequencies.  The correlation between GX 17+2's radio and IR emission suggests the presence of a synchrotron jet.

The lack of a periodicity in the RXTE ASM soft X-ray data can be explained if the jet is responsible for only a small amount of the soft X-ray flux.  The periodic signal would then be lost in the stronger X-ray emission from the accretion process.  That GX 17+2 was detected at a flux of $\sim$10 mJy at 24 $\mu$m supports this interpretation, for neither donor star nor accretion disk would contribute to the emission at that wavelength.  For example, STAR-PET\footnote{http://ssc.spitzer.caltech.edu/tools/starpet/} predicts a 24 $\mu$m flux of 0.011 mJy for NP Ser.

Assuming our interpretation of a jet is correct, one could ask whether this periodic IR activity represents the orbital period of an eccentric binary, with the jet turning on at periastron as used to explain Cir X-1, or whether it represents the precession period of the jet.  However, periastron passage implies an eccentric orbit.  GX 17+2 is a persistent X-ray source, as are all the Z sources.  Continuous accretion by Roche lobe overflow requires these systems to have circular orbits.  Therefore, we believe our observations support the possibility of GX 17+2 having a precessing synchrotron jet, making GX 17+2 an obvious choice for a future simultaneous multi-wavelength studies.

\acknowledgements 
This material is based upon work supported by the National Aeronautics and Space Administration under Proposal No. 92042 issued through the Science Mission Directorate.  Support for this work was also provided by the New Mexico Space Grant Consortium, and the New Mexico Higher Education Department.  Results provided by the ASM/RXTE teams at MIT and at the RXTE SOF and GOF at NASA's GSFC, by the National Optical Astronomy Observatory (NOAO)/Association of Universities for Research in Astronomy (AURA)/National Science Foundation (NSF), and based on observations obtained with the Apache Point Observatory 3.5-meter telescope, which is owned and operated by the Astrophysical Research Consortium.  This research has made use of the NASA/ IPAC Infrared Science Archive, which is operated by the Jet Propulsion Laboratory, California Institute of Technology, under contract with the National Aeronautics and Space Administration.

{\it Facilities:} \facility{RXTE}, \facility{CTIO:1.3m}, \facility{KPNO:2.1m}, \facility{APO:3.5m}, \facility{VLA}.

\bibliography{biblio2}

\begin{deluxetable}{lccccl}
\tablecolumns{6}
\tablewidth{0pt}
\tablecaption{GX 17+2 {\it K}-Band Bright Episode Midpoint Times}
\tablehead{\colhead{Date} & \colhead{JD} & \colhead{Hel.Cor. (days)} & \colhead{Peak {\it K} mag\tablenotemark{a}} & \colhead{Duration (hours)} & \colhead{Site}}
\startdata
1999-06-26\tablenotemark{1} & 2451355.9740 & 0.0058 & 13.98 &Single & Keck\\
2006-07-12 & 2453928.7250 & 0.0056 & 13.91 & 2.5 &APO\\ 
2007-05-27 & 2454247.9333 & 0.0051 & 14.01 & 3.2 &KPNO\\
2007-05-30 & 2454250.9375 & 0.0052 & 13.83 & 4.2 &KPNO\\
2007-10-04 & 2454377.5875 &$-$0.0006 & 14.05 & 0.2 &APO\\
2008-07-13 & 2454660.7983 & 0.0056 & 13.91 & 2.8 &KPNO\\
2008-10-21\tablenotemark{2} & 2454760.7592 &$-$0.0023 &  -    & Single &{\it Spitzer}\\
2009-07-01 & 2455013.7484 & 0.0057& 14.30 & 2.5 & APO\\
\enddata
\tablenotetext{a}{Magnitudes reported are of the blend of GX 17+2 and NP Ser.}
\tablenotetext{1}{\cite{Callanan}}
\tablenotetext{2}{\cite{Wachter}}
\label{bright_events}
\end{deluxetable}

\begin{deluxetable}{lcccl}
\tablecolumns{4}
\tablewidth{0pt}
\tablecaption{GX 17+2 Observation Information}
\tablehead{\colhead{Date} & \colhead{Duration (hours)} & \colhead{IR State} &\colhead{Instrument} & \colhead{Site}}
\startdata
2006-07-01 & 2.5 & Faint & NICFPS & APO\\
2006-07-12 & 2.5 & Bright & NICFPS & APO\\
2006-07-14 & 2.5 & Faint & NICFPS & APO\\
2007-05-25 & 4.2 & Faint & SQIID & KPNO\\
2007-05-26 & 4.2 & Faint & SQIID & KPNO\\
2007-05-27 & 4.2 & Bright &SQIID & KPNO\\
2007-05-28 & 4.2 & Faint & SQIID & KPNO\\
2007-05-29 & 4.2 & Faint & SQIID & KPNO\\
2007-05-30 & 4.2 & Bright & SQIID & KPNO\\
2007-10-04 & 0.2 & Bright & CorMASS & APO\\
2008-07-08 & 2.8 & Faint & SQIID & KPNO\\
2008-07-13 & 2.8 & Bright & SQIID & KPNO\\
2008-10-21 & Single & Bright & IRAC & {\it Spitzer}\\
2009-07-01 & 2.5 & Bright & NICFPS& APO\\
\enddata
\label{observations}
\end{deluxetable}

\begin{figure}[hb]
\begin{center}
\includegraphics[scale=0.7]{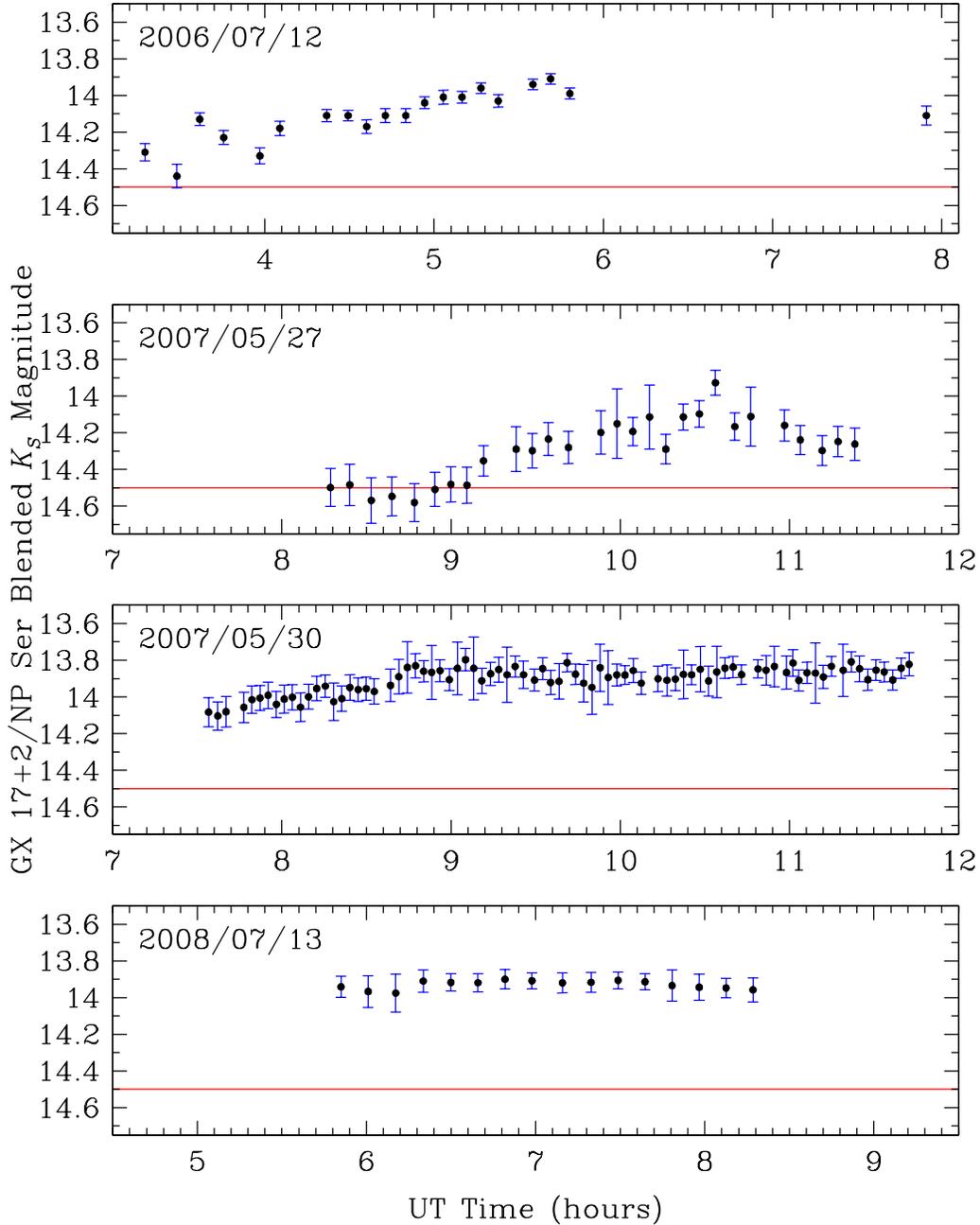}
\caption{{\it K}-band light curves for the blend of NP Ser and GX 17+2 are shown with 1$\sigma$ error bars.  The blend is shown clearly brighter than the {\it K} magnitude of NP Ser as measured by \cite{Callanan}, indicated with a solid line in each panel.  The shapes and peak levels of an IR bright episode vary, making a unifying folded light curve impractical.}
\label{lightcurves}
\end{center}
\end{figure}

\begin{figure}[hb]
\centering
\includegraphics[scale=0.5]{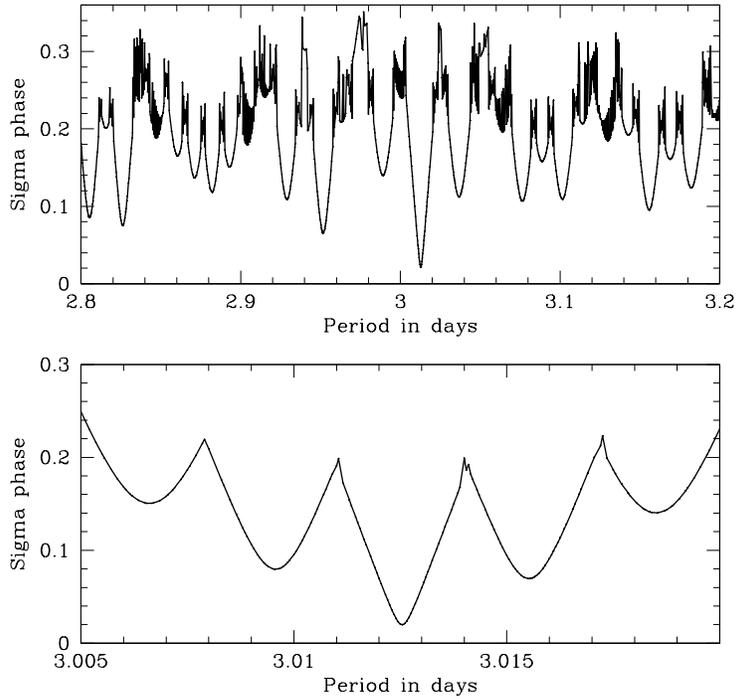}
\caption{Period vs. the dispersion in phase for the times when GX 17+2 was IR bright. The top figure includes bright episode times listed in Table 1 between 2006 and 2008. The bottom figure includes the bright episode reported by \cite{Callanan} for June 1999. The minimum shown in the bottom figure corresponds to the period when the dispersion in the bright time phases was at its minimum. This occurs at a period of 3.01245 days.  The multi-cusp pattern in the bottom figure (and present but not visible in the top figure) is due to our data having been taken at nearly the same time each year.}
\label{period}
\end{figure}

\end{document}